# Neutron Diffraction Reveals the Existence of Confined Water in Triangular and Hexagonal Channels of Modified YPO$_4$ at Elevated Temperatures

S. K. Mishra,[†] R. S. Ningthoujam,[#,§] R. Mittal,[†,#] R. K. Vatsa,[#§] M. Zbiri,[¥] K. Shitaljit Sharma,[§] B. P. Singh,[§] P. U. Sastry,[†,#] T. Hansen,[¥] H. Schober,[¥] and S. L. Chaplot[†,#]

[†]Solid State Physics Division, Bhabha Atomic Research Centre, Mumbai 400085, India
[#]Homi Bhabha National Institute, Anushaktinagar, Mumbai 400094, India
[§]Chemistry Division, Bhabha Atomic Research Centre, Mumbai 400085, India
[¥]Institut Laue-Langevin, 71 avenue des Martyrs, Grenoble Cedex 9, 38042, France

We provide experimental evidence for confinement of water molecules in the pores of hexagonal structure of YPO$_4$ at elevated temperatures upto 600 K using powder neutron diffraction. In order to avoid the large incoherent scattering from the hydrogen, deuterated samples of doped YPO$_4$:Ce-Eu were used for diffraction measurements. The presence of water molecules in the triangular and hexagonal pores in the hexagonal structure was established by detailed simulation of the diffraction pattern and Rietveld refinement of the experimental data. It was observed that the presence of water leads specifically to suppression of the intensity of a peak around $Q$ = 1.04 Å$^{-1}$ while the intensity of peaks around $Q$=1.83Å$^{-1}$ is enhanced in the neutron diffraction pattern. We estimate the number of water molecules as 2.36 (6) per formula units at 300 K and the sizes of the hexagonal and triangular pores as 7.2 (1) Å and 4.5 (1) Å, respectively. With increase in temperature, the water content in both the pores decreases above 450 K and vanishes around 600 K. Analysis of the powder diffraction data reveals that the hexagonal structure with the pores persist up to 1273 K, and transforms to another structure at 1323 K. The high temperature phase is not found to have the zircon or the monazite type structure, but a monoclinic structure (space group $P2/m$) with lattice parameters $a_m$= 6.826 (4) Å, $b_m$= 6.645 (4) Å, $c_m$= 10.435 (9) Å, and $β$= 107.21 (6)°. The monoclinic structure has about 14 % smaller volume than the hexagonal structure which essentially reflects the collapse of the pores. The phase transition and the change in the volume are also confirmed by x-ray diffraction measurements. The hexagonal to the monoclinic phase transition is found to be irreversible on cooling to room temperature.

**Introduction**

Rare-earth (RE) phosphates and its derivatives (REPO$_4$:RE=La, Ce, Gd or Y) exhibit exotic properties like chemical stability, excellent high-temperature properties, high radiation damage tolerance, high luminescence quantum yield, high energy band gap, etc[1-12]. Due to a sharp emission, these compounds are also used as host for luminescence applications[6, 7, 13] including lasers and display devices. A combination of rare-earth phosphates and silicates is used as host for nuclear wastes[14]. At ambient condition, depending on the ionic radii of RE ion and synthesis route, the compounds can crystalize in tetragonal, hexagonal and monoclinic phases[13, 15-18]. A smaller ionic radius compared to the ionic radius of Gd generally adopts the tetragonal structure, whereas other orthophosphates have the lower-symmetry monoclinic structure. Depending on the synthesis conditions, GdPO$_4$, TbPO$_4$, DyPO$_4$, and HoPO$_4$ can adopt either zircon or monazite structure[13, 16, 19-21].

LaPO$_4$ crystallizes in two polymorphs, namely, the tetragonal phase (xenotime structure) and the monoclinic phase (monazite structure). Yttrium orthophosphate, YPO$_4$, also has the tetragonal symmetry (xenotime type)[6]. Doping YPO$_4$ matrix with Ce$^{3+}$ and Eu$^{3+}$, enhances luminescence intensity and scintillation efficiency which makes it practically suitable candidate in scintillators industry[6, 22-25]. Further, dopedYPO$_4$remains in the tetragonal phase. However, on increasing the concentration of Eu$^{3+}$ in Ce$^{3+}$ doped YPO$_4$, a tetragonal to hexagonal phase transition occurs[6]. The hexagonal phase ofYPO$_4$:Eu:Ce has a zeolite like configuration, in which many pores are available along the *c*-axis (like a channel) and water can be confined in these pores. Recently, we found that the hexagonal phase of modified REPO$_4$exhibits a high temperature stability[6]. Thermogravimetric analysis-mass spectrometry (TGA-MS) measurements show that the hexagonal phase of modified REPO$_4$ (RE=Y, La) can retain water molecules up to very high temperatures (~ 1073 K). Upon heating above 1073 K, the hexagonal phase transforms to a tetragonal phase due to the removal of water. It is well known that the hexagonal phase is stable because water molecules are occupying the pores along c-axis. This was found in Ce/Bi co-doped YPO$_4$:Eu. Presence of water persists up to 800 $^o$C (1073 K).The confined water molecules are not frozen even at 215 K[26]. In order to understand the anomalous behavior of confined water, we need to identify the positions/locations of water molecules.

In X-ray diffraction, scattering is due to interaction of X-rays with the electronic charge cloud of atoms, whereas in neutron diffraction, scattering is due to interaction of neutron with nuclei of atoms. Hence, neutron diffraction has an advantage over X-ray diffraction in terms of detection of lighter atoms in presence of heavy atoms. This makes neutron diffraction techniques suitable to identify the existence of confined water in pores. In this paper, we present detailed structural studies as a function of temperature to investigate the pore sizes, nature of bonding of water molecules with host lattice and the structural phase transition. Interestingly, water molecules occupy the triangular and hexagonal channels of hexagonal phase REPO$_4$, which, to the best of our knowledge, is confirmed by the present neutron diffraction studies. Also, a new monoclinic phase with space group *P2/m* is reported at high temperatures.

**Experimental**

15at.% Ce and 5 at.% Eu doped YPO$_4$ (YPO$_4$:Ce-Eu) powders were prepared by taking precursors such as Y$_2$O$_3$, Ce(CH$_3$COO)$_3$, Eu(CH$_3$COO)$_3$ and NH$_4$H$_2$PO$_4$[reference6]. The metal precursors (Y$_2$O$_3$, Ce(CH$_3$COO)$_3$, Eu(CH$_3$COO)$_3$) were dissolved in HNO$_3$ acid in a round bottom flask (RBF). The excess acid was removed by alternate warming and cooling processes with addition of D$_2$O in Ar atmosphere. Stoichiometric ratio of (NH$_4$)H$_2$PO$_4$ was dissolved in D$_2$O and solution was injected to RBF containing metal ions precursors. In order to avoid H or water present on the surface of particles, D$_2$O was added at different time intervals during heating at 100-160°C in Ar atmosphere. About 2-3 liters of D$_2$O were consumed to get 10-15 g of YPO$_4$:Ce-Eu powder. Dried powder was transferred into glass bottle and sealed under Ar atmosphere using glove box. The obtained powder sample is considered as deuterated sample. Powder X-ray diffraction was used to characterize the sample.

Temperature dependent neutron powder diffraction experiments were performed at the high-flux D20 diffractometer[27] at the Institut Laue- Langevin, Grenoble, France in temperature range 300-1323 K. The high resolution mode (take-off angle of 120°) was selected with a wavelength of 1.3594 Å. The measurements were performed using the deuterated sample, since the usual hydrogenous sample gives a large incoherent background in the neutron diffraction data.

At each temperature, the crystal structures were analyzed using the Rietveld refinement program FULLPROF[28]. A Thompson-Cox-Hastings pseudo-Voigt with axial divergence asymmetry function was used to model the peak profiles. The background was fitted using a sixth order polynomial. Except for the occupancy parameters of the atoms, which were fixed corresponding to the nominal composition, all other parameters., i.e., scale factor, zero displacement, structural parameters were refined. In all the refinements, the data over full angular range has been used, although in the figures only a limited range has been shown to highlight the changes. To complement the results from neutron diffraction, powder x-ray diffraction experiments were also performed at selected temperatures.

**Results and discussion**

Figure 1 (a) shows the result of Rietveld refinement of room-temperature powder x-ray diffraction data using the hexagonal structure of $YPO_4$ with empty pores. It is evident that all the reflections except the marked ones are accounted for using the hexagonal structure. This confirms the formation of the hexagonal phase at room temperature. We have explored various possibilities and found that the additional reflections marked with arrows correspond to unreacted ammonium deuterium phosphate and cerium acetate. It is important to notice that these additional peaks disappear on heating at high temperature (~473 K) (discussed later). Figure 1 (b) shows the comparison of powder X-ray and neutron diffraction patterns of the sample (intensity as a function of $Q$, $Q = 4\pi(sin\theta)/\lambda$). It is evident from this figure that the neutron diffraction data also have additional peaks as observed in X-ray diffraction data. We have refined both the x-ray and the neutron diffraction data using the hexagonal phase. It is important to notice that the refined structure is not able to account the intensity of certain reflections around $Q=$ 1.04 and 1.83 Å$^{-1}$ (Fig. 1 (c)).

It is well known that the hexagonal phase has a zeolite configuration, with many pores along the $c$-axis. Pore size around 5-6 Å is sufficiently large for $H_2O/D_2O$ molecules to occupy the pores. This was found in Ce/Bi co-doped YPO4:Eu. Based on the TGA-MS and NMR studies, the presence of water molecules inside the pores has been reported (Reference 26).

To investigate the effect of confined water on neutron diffraction data, we simulated the diffraction patterns assuming various possible ways the water molecules could be contained in the pore structure (Figure 2). It may be noted that in view of the large incoherent background from the hydrogen, deuterated samples of modified $YPO_4$:Ce-Eu were used for diffraction measurements. In simulation, we considered the neutron scattering length of deuterated water molecules. In the simulation as well as in the Rietveld refinement of the neutron diffraction data described below, we assume the $D_2O$ as a point particle with the total neutron coherent scattering length of $D_2O$. The Debye-Waller factor of $D_2O$ includes both the size of the $D_2O$ molecules and the thermal vibrational amplitude. The simulation result shows that in pure hexagonal phase (without water) the reflection appearing around $2\theta= 12.9°$ is the strongest one (figure 2 top left panel). As it can be seen from the figure, the hexagonal structure has two type of pores, namely, the triangular ring (confined in triangular cage) and the hexagonal ring (confined in hexagonal cage). The terms pore, ring and cage are often used interchangeably in the literature. Assuming that water is confined only in the triangular ring, the simulated diffraction pattern shows that the intensity of all reflections decreases except for the reflection at 12.9 degree (figure 2 top right panel). On the other hand, if water is confined only in the hexagonal cage, the intensity of the first peak decreases while the peak intensity appearing at around $2\theta= 22.5°$ increases (figure 2). Finally, we considered deuterated water molecules in both the rings and find substantially different diffraction patterns as shown in figure 2.

Now, if we recall the neutron diffraction pattern at room temperature, we see that the measured intensity of first peak around $2\theta=12.9°$ is suppressed while the intensity of peaks around $22.5°$ is enhanced with respect to the calculated diffraction pattern for the pure hexagonal phase without water. Thus, we again refine the neutron diffraction data using the model with the water molecules in both the cages. We found that the insertion of water molecules in pores modifies the intensities (Figure 3 (a)). Consequently, the fitting in figure 3 has improved as compared to that in figure 1, and the refined intensities of peaks match well with experiments. This might be regarded as a signature of the presence of water in the cages. The structural parameter of $YPO_4$:Ce-Eu.$nD_2O$ obtained by Rietveld refinement of the neutron diffraction data at 300 K using the hexagonal structure are shown in Table 1. Occupancies of $D_2O$ molecules in the triangular and hexagonal rings are found to be 0.37 and 0.22 respectively. Total number of $D_2O$ molecules in unit cell is $(0.37+0.22)\times12= 7.08$, which correspond to 2.36 (6) $D_2O$ molecules per formula unit of $YPO_4$. The hexagonal and triangular pore sizes are estimated as 7.2 (1) Å and 4.5 (1) Å, respectively. On increasing the temperature, the pore sizes increase as a result of increment in lattice parameters. The structure of $YPO_4$:Ce-Eu.$nD_2O$ obtained by Rietveld refinement of neutron diffraction at 300 K using the hexagonal structure is shown in figure 3 (b).

Further, to identify the temperature where the amount of confined water vanishes, we carried out temperature dependent neutron diffraction study. Figure 3 (c) shows a part of the neutron powder diffraction pattern of $YPO_4$:Ce-Eu as a function of temperature in the range 300–573 K. It is observed that as temperature increases from room temperature, the intensity of certain peaks (due to ammonium hydrogen phosphate and cerium acetate) decreases and finally vanishes around 573 K. The intensity of peaks around $Q = 1.04$ Å$^{-1}$ ($2\theta =12.9°$) increases with increasing temperature, while intensity of peaks around $Q=1.83$Å$^{-1}$ ($2\theta =22.5°$) decreases below 550 K. The integrated intensity of these two peaks is nearly temperature independent in the temperature range 600 – 1200 K, and decreases with further increase of temperature (figure 4(a)). The ratio of the integrated intensities of these characteristic peaks is shown in figure 4(b). It is found that the ratio of these peaks increases with increasing temperature, exhibiting an anomaly around 550 K. Heating above 550 K, the ratio becomes nearly temperature independent upto 1200 K, and further decreases sharply to finally vanishes at 1323K. Figure 4(c) depicts the variation of the amount of confined water obtained using Rietveld refinement of the powder neutron diffraction data. We found that as the temperature increases the water content in both the pores decreases above 450 K, and vanishes around 600 K.

Figure 5 (a) shows a part of the neutron powder diffraction pattern of YPO$_4$:Ce-Eu as a function of temperature in the range 573–1373 K. The powder neutron diffraction patterns also show dramatic changes as a function of temperature, especially in terms of dissimilar broadening, intensity and splitting of various peaks. We also noticed the disappearance and appearance of additional reflection above 1223 K. At 573 K, all the Bragg reflections present in the powder diffraction patterns can be indexed with the hexagonal phase, and the refinement without the water molecules leaves some excess intensity at the peak round 12.9° (see figure 5(b)). However, including the water molecules in pores results in increasing the intensity of this peak, which matches very well the observed data (see Figure 5(c)). The neutron data at and above 723 K could be explained with the hexagonal phase without any water. This suggests that above 723 K, water molecules are not detected in our neutron diffraction study. Rietveld refinement of the powder neutron diffraction data shows that the temperature dependent neutron diffraction patterns could be indexed using the hexagonal structure without any water from 723 K to 1223 K. At a higher temperature, 1323 K, the refinement of the powder neutron diffraction data using the hexagonal phase is not satisfactory, suggesting a different structure.

Based on high temperature X-ray diffraction study, Luwang et al[6] reported that on heating above 1223 K, the sample undergoes a hexagonal to zircon type tetragonal phase transition. In view of this, we refined the neutron diffraction data collected at 1323 K using the zircon type tetragonal structure as shown in figure 6 (a). But the tetragonal phase could not account for all the reflections present in the powder neutron diffraction data at 1323K, which suggests the incorrectness of the structure. We have also tried other possible structures as reported in the literature like the monoclinic structure of LaPO$_4$ (monazite structure), but the attempt was unsuccessful as the peak near 20° could not be indexed as shown in figure 6 (b). Further, we have explored various possibilities and found that another monoclinic structure (with space group *P2/m*) with lattice parameters $a_m$= 6.826 (4) Å, $b_m$= 6.645 (4) Å, $c_m$= 10.435 (9) Å, and $β$= 107.21 (1) ° is able to index (Figure 6 (c)) all the peaks present in the neutron diffraction pattern at 1323 K. It is important to notice the that there is 14% volume drop across the hexagonal to the monoclinic phase transition at around 1323 K; (as the volume of the hexagonal (z=3) and the monoclinic (z=6) phases are 262.6 and 452.2 Å$^3$ respectively). This change in the volume can be expected from the collapse of all the empty pores in the hexagonal structure. In fact, the volume of the monoclinic structure matched close to the volume of the unmodified YPO$_4$ that does not have any pores of the zeolite-like hexagonal structure.

Figure 7 depicts the variation of lattice parameters as a function of temperature in the stability region of the hexagonal phase. The lattice parameters increase with increasing temperature and do not show any anomaly in the entire explored temperature range of 573- 1273 K. The thermal expansion coefficient along the hexagonal *c*-axis is larger compared to that along the *a*-axis. The average coefficient of thermal expansion of YPO$_4$:Ce-Eu along the a- and c-axes of the hexagonal phase were found to have values of 3.2 (1) ×10$^{-6}$ K$^{-1}$, and 8.5 (2) ×10$^{-6}$ K$^{-1}$ respectively. The difference among these values indicates the anisotropy in the thermal expansion. The coefficient of volume thermal expansion is found to be 14.9(3)×10$^{-6}$ K$^{-1}$.

A portion of the powder X-ray diffraction pattern of YPO$_4$:Ce-Eu at selected temperatures is shown in Figure 8(a). From 573 to 1223 K, all the Bragg reflections present in the powder diffraction patterns could be indexed with the hexagonal phase. The powder X-ray diffraction patterns reveal the appearance of additional reflections above 1223 K which could be ascribed to the monoclinic phase found above from the neutron diffraction. Presence of the additional characteristic peaks of the monoclinic phase at 1323 K clearly provides an evidence for phase coexistence and confirms the first order nature of the hexagonal to monoclinic phase transition. We found that the refinement of the powder diffraction data at 1483 K was unsuccessful using the hexagonal phase, or the zircon type tetragonal phase as well as the monazite type monoclinic phase, since a number of peaks present in powder diffraction data could not be indexed (see figure 8(b & c)). However, as shown earlier in case of the powder neutron diffraction, another monoclinic phase (with space group P2/m) is able to index all the reflections(see figure 8(d).The refined lattice parameters at 1483 K are $a_m$= 6.832 (4) Å, $b_m$= 6.653 (4) Å, $c_m$= 10.429 (9) Å, and β = 107.20 (1)°. These values from the x-ray diffraction data are consistent with the values found above from the neutron diffraction data at 1323 K, which gives confidence in the correctness of the new monoclinic structure determined here.

Remarkably, we notice that the small difference in the phase transition temperature observed in the neutron and X-ray diffraction cases is related to experimental history, namely, the time elapsed and rate of heating conditions. It is also important to notice that the sample remains in the monoclinic phase (high temperature phase) after cooling to room temperature, which suggests that the hexagonal to the monoclinic phase transition is irreversible in nature.

**Conclusions**

In summary, we have carried out high temperature powder neutron and X-ray diffraction studies on YPO$_4$:Ce-Eu. We provide experimental evidence for confinement of water molecules in the pores of the hexagonal structure at elevated temperatures up to 600 K. At room temperature, we found about 1.48 (4) and 0.88 (4) water molecules per formula unit of YPO4 in the triangular and hexagonal pores respectively. Rietveld refinement of the powder diffraction data revealed that the diffraction patterns could be indexed using the hexagonal structure up to 1273 K. Careful inspection of the temperature dependence of the diffraction data above 1273 K indicated a structural phase transition. Bragg reflections present in the high temperature neutron diffraction pattern at 1323 K could not be indexed using neither zircon type tetragonal phase nor the monazite type monoclinic phase, as reported in literature. We found that the monoclinic structure (space group *P2/m*) with lattice parameters $a_m$= 6.826 (4) Å, $b_m$= 6.645 (4) Å, $c_m$= 10.435 (9) Å, and *β*= 107.21 (6)° (volume V= 452.23(1) Å$^3$)leads to indexing of all the peaks present in neutron diffraction patterns at 1323 K. It is also important to note that the samples remain in the monoclinic phase (high temperature phase), even after cooling to room temperature, suggesting that the hexagonal to the monoclinic phase transition is irreversible in nature.

**Acknowledgements**

S. L. Chaplot would like to thank the Department of Atomic Energy, India for the award of Raja Ramanna Fellowship.

1   A. al-Wahish, U. al-Binni, C. A. Bridges, S. Tang, Z. Bi, M. P. Paranthaman, A. Huq, and D. Mandrus, Chemistry of Materials **28**, 7232 (2016).
2   N. Sheng, et al., Journal of the American Chemical Society **138**, 6171 (2016).
3   L. Li, Y. Wang, B.-H. Lei, S. Han, Z. Yang, K. R. Poeppelmeier, and S. Pan, Journal of the American Chemical Society **138**, 9101 (2016).
4   M.-J. Sie, C.-H. Lin, and S.-L. Wang, Journal of the American Chemical Society **138**, 6719 (2016).
5   F. G. Alabarse, J. Rouquette, B. Coasne, A. Haidoux, C. Paulmann, O. Cambon, and J. Haines, Journal of the American Chemical Society **137**, 584 (2015).
6   M. N. Luwang, R. S. Ningthoujam, Jagannath, S. K. Srivastava, and R. K. Vatsa, Journal of the American Chemical Society **132**, 2759 (2010).
7   A. A. Kaminskii, M. Bettinelli, A. Speghini, H. Rhee, H. J. Eichler, and G. Mariotto, Laser Physics Letters **5**, 367 (2008).
8   U. K. a. D. Holtstam, European Journal of Mineralogy, **16**, 117 (2004).
9   J. B. Davis, D. B. Marshall, and P. E. D. Morgan, Journal of the European Ceramic Society **20**, 583 (2000).
10  Y. Hikichi, Ota, T., Daimon, K., Hattori, T. and Mizuno, M., Journal of the American Ceramic Society **81**, 2216 (1998).
11  J. B. Davis, Marshall, D. B., Housley, R. M. and Morgan, P. E. D, Journal of the American Ceramic Society **81**, 2169 (1998).
12  A. Meldrum, L. A. Boatner, and R. C. Ewing, Physical Review B **56**, 13805 (1997).
13  R. Lacomba-Perales, D. Errandonea, Y. Meng, and M. Bettinelli, Physical Review B **81**, 064113 (2010).
14  H. Li, S. Zhang, S. Zhou, and X. Cao, Inorganic Chemistry **48**, 4542 (2009).
15  D. E. a. F. J. Manjon, Prog. Mater. Sci. **53**, 711 (2008).
16  M. Giarola, A. Sanson, A. Rahman, G. Mariotto, M. Bettinelli, A. Speghini, and E. Cazzanelli, Physical Review B **83**, 224302 (2011).
17  L. A. Boatner, Reviews in Mineralogy and Geochemistry **48**, 87 (2002).
18  F. X. Zhang, J. W. Wang, M. Lang, J. M. Zhang, R. C. Ewing, and L. A. Boatner, Physical Review B **80**, 184114 (2009).
19  J. Ruiz-Fuertes, A. Hirsch, A. Friedrich, B. Winkler, L. Bayarjargal, W. Morgenroth, L. Peters, G. Roth, and V. Milman, Physical Review B **94**, 134109 (2016).
20  E. Stavrou, A. Tatsi, C. Raptis, I. Efthimiopoulos, K. Syassen, A. Muñoz, P. Rodríguez-Hernández, J. López-Solano, and M. Hanfland, Physical Review B **85**, 024117 (2012).
21  A. H. Krumpel, A. J. J. Bos, A. Bessière, E. van der Kolk, and P. Dorenbos, Physical Review B **80**, 085103 (2009).
22  F. Angiuli, E. Cavalli, P. Boutinaud, and R. Mahiou, Journal of Luminescence **135**, 239 (2013).
23  G. Blasse and B. C. Grabmaier, *Luminescent Materials*, 1994).
24  J. Chen, Q. Meng, P. S. May, M. T. Berry, and C. Lin, The Journal of Physical Chemistry C **117**, 5953 (2013).
25  M. N. Luwang, R. S. Ningthoujam, S. K. Srivastava, and R. K. Vatsa, Journal of the American Chemical Society **133**, 2998 (2011).
26  R. S. Ningthoujam, Pramana – J. Phys., **80**, 1055 (2013).
27  C. H. Thomas, F. H. Paul, E. F. Henry, T. Jacques, and C. Pierre, Measurement Science and Technology **19**, 034001 (2008).
28  R.-C. J, Physica B **192**, 55 (1993).

**TABLE I.** Structural parameters of YPO$_4$:Ce-Eu.nD$_2$O obtained by Rietveld refinement of neutron diffraction at 300 K, using a hexagonal structure with space group ***P6222***. In Rietveld refinement, D$_2$O was assumed as a molecule with its total neutron coherent scattering length. Occupancies of D$_2$O molecules for the triangular and hexagonal rings are found to be 0.37 (1) and 0.22 (1) respectively. Total number of D$_2$O molecules in unit cell is (0.37+0.22)×12= 7.08, which corresponds to 2.36 (6) water molecules per formula unit.

|  | Positional coordinates | | | Thermal parameter |
| --- | --- | --- | --- | --- |
| **Atoms** | X | Y | Z | B (Å)$^2$ |
| **Y/Ce/Eu** | 0.5 | 0 | 0 | 0.15(6) |
| **P** | 0.5 | 0 | 0.5 | 2.24 (9) |
| **O** | 0.431(7) | 0.134(3) | 0.359(2) | 0.69 (9) |
| **D$_2$O (triangular)** | 0.643 (2) | 0.245(1) | 0.665(3) | 6.50 (2) |
| **D$_2$O (hexagonal)** | 0.007(3) | 0.066(1) | 0.953(4) | 2.65 (7) |
|  |  |  |  |  |
| **Lattice parameters (Å)** | | | | |
| a =b= 6.897 (5) (Å), c= 6.3089 (3) (Å) | | | | |
| R$_p$= 5.48; R$_{wp}$= 8.31; R$_{exp}$= 4.23; χ$^2$ = 3.86 | | | | |

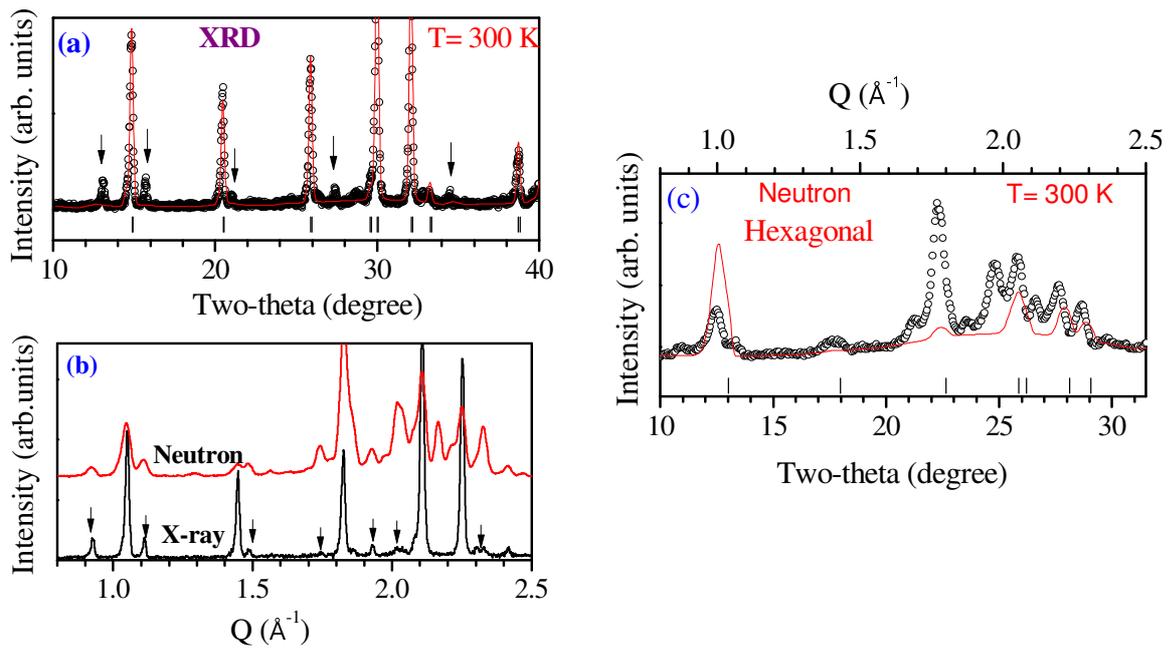

**Figure 1.** Observed (circle) and calculated (line) profiles obtained after the Rietveld refinement of YPO$_4$:Ce-Eu, using a hexagonal structure of (a) X-ray diffraction and (c) neutron diffraction data at ambient temperature. Comparison of the powder X-ray and neutron diffraction patterns is shown in figure (b).

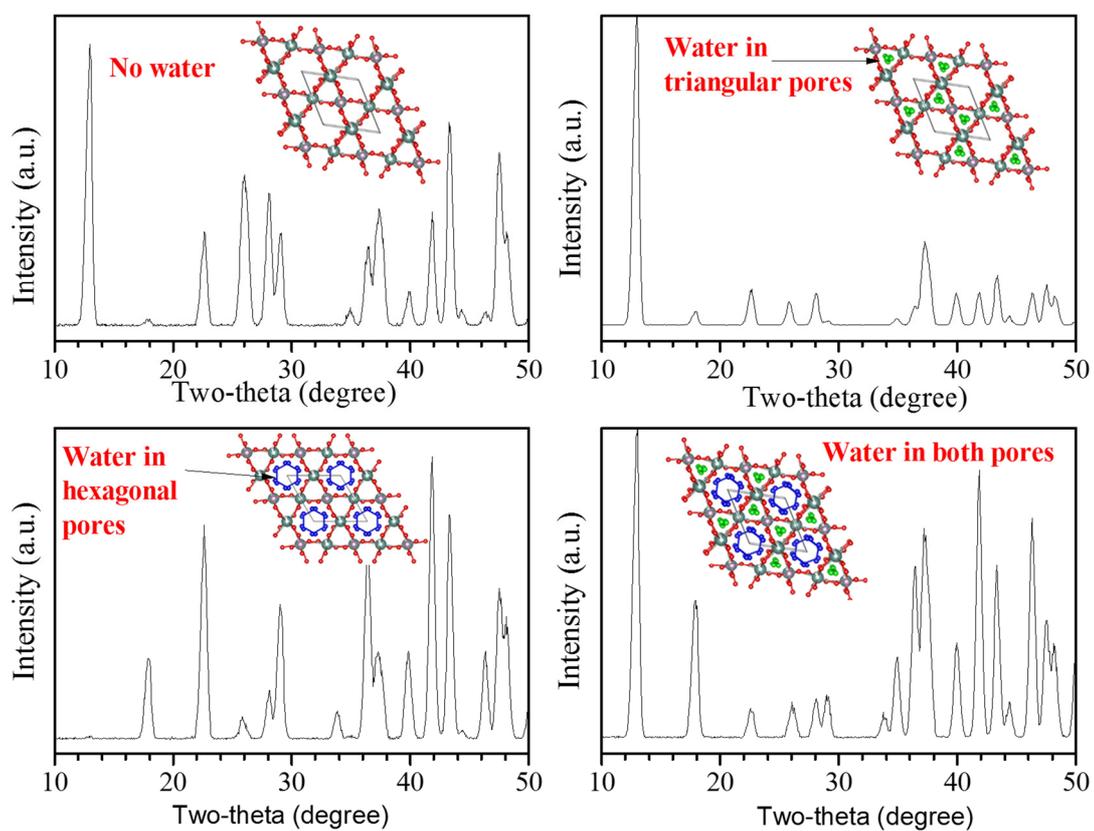

**Figure 2.** Simulated powder neutron diffraction patterns of YPO4:Ce-Eu, with a hexagonal structure, which acts as host in which water ($D_2O$) molecules occupy the pores at 300 K.

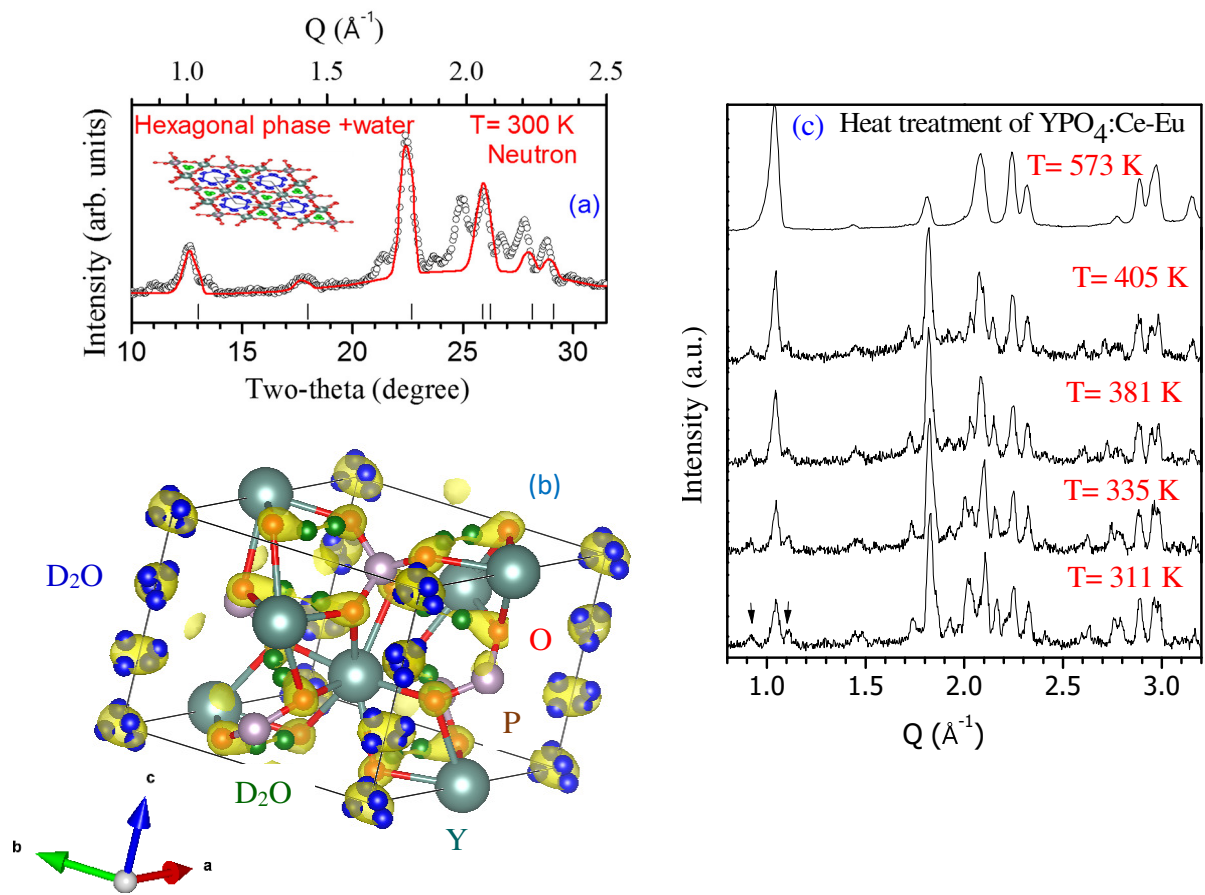

**Figure 3.** (a)Observed (circle) and calculated (line) neutron diffraction profiles obtained after the Rietveld refinement of YPO$_4$:Ce-Eu, using a hexagonal symmetry with water molecules in pores at 300 K. The structure is shown in (b). The hexagonal and triangular pore sizes are estimated as 7.2 (1) Å and 4.5(1) Å, respectively. We found 7.08D$_2$O molecules per unit cell. Evolution of a portion of the powder neutron diffraction pattern of YPO$_4$ at selected temperatures is shown in (c).

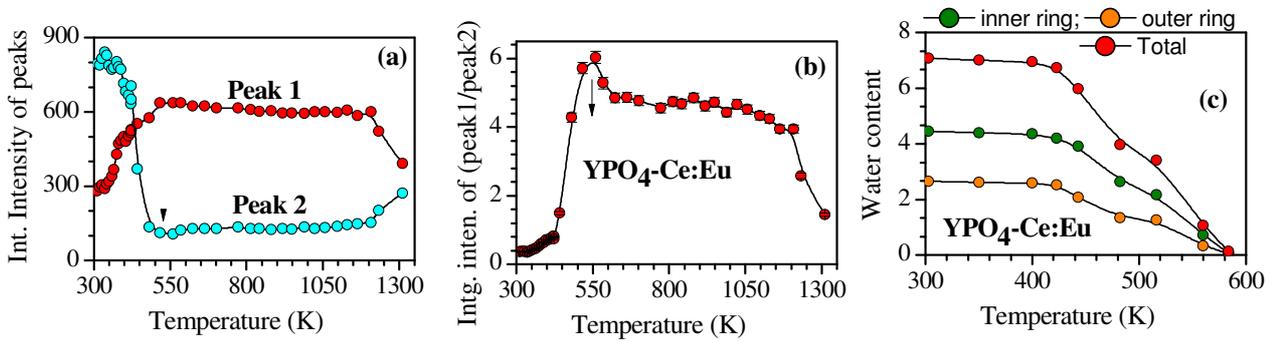

**Figure 4.** Variation of (a) the integrated intensity of characteristic peaks around $Q = 1.04 \text{Å}^{-1}$ ($2\theta = 12.9°$: peak 1) and $Q = 1.83 \text{Å}^{-1}$ ($2\theta = 22.5°$: peak 2); (b) ratio of the two peak intensities and (c) content of water with temperature.

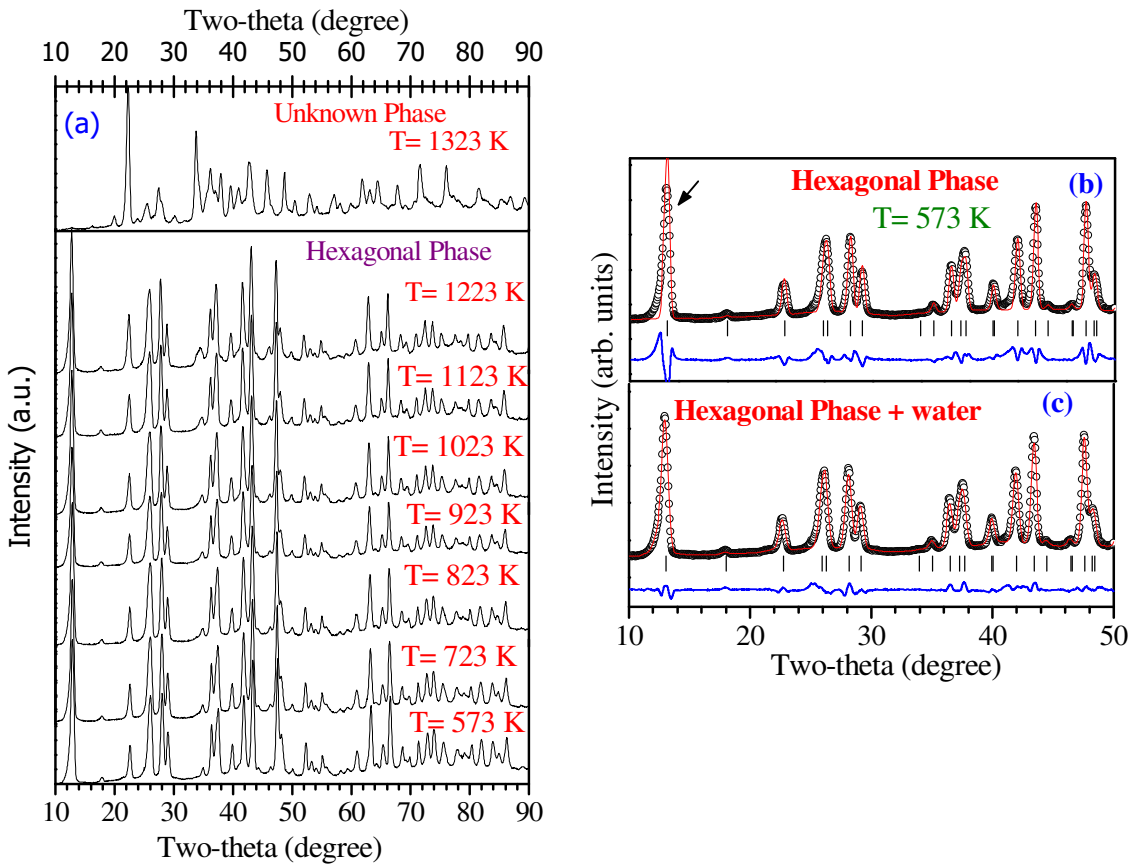

**Figure 5** (a) Evolution of a portion of powder neutron diffraction patterns of YPO$_4$:Ce-Eu at selected temperatures. Observed (circle) and calculated (line) profiles obtained after the Rietveld refinement of YPO$_4$:Ce-Eu using a hexagonal structure (b) without and (c) with water at 573 K.

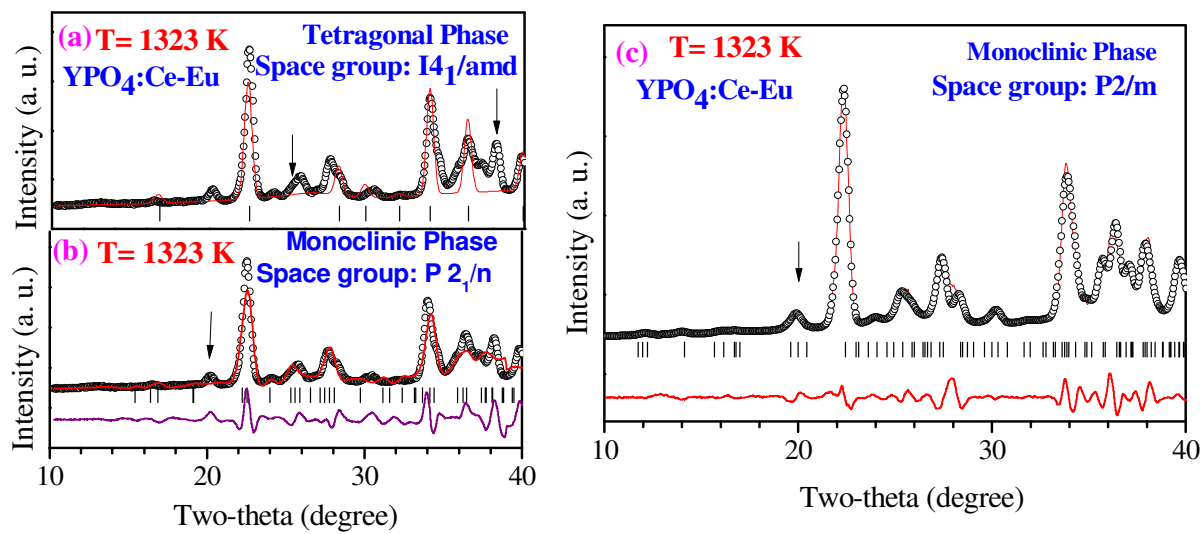

**Figure 6** Observed (circle) and calculated (line) neutron diffraction profiles obtained after the Rietveld refinement (structure-free Lebail fitting) of YPO$_4$:Ce-Eu using (a) tetragonal (zircon), (b) monoclinic (monazite with space group:P2$_1$/n) and(c) monoclinic with space group: P2/m at 1323 K, respectively.

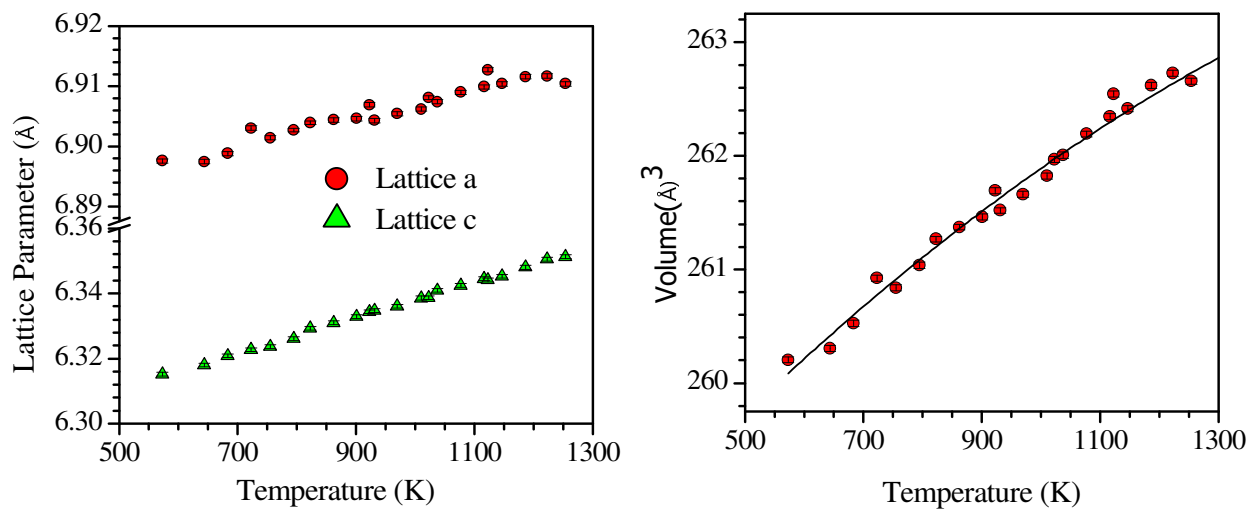

**Figure 7.** Temperature evolution of the lattice parameters and volume, obtained from the Rietveld refinements of neutron diffraction data.

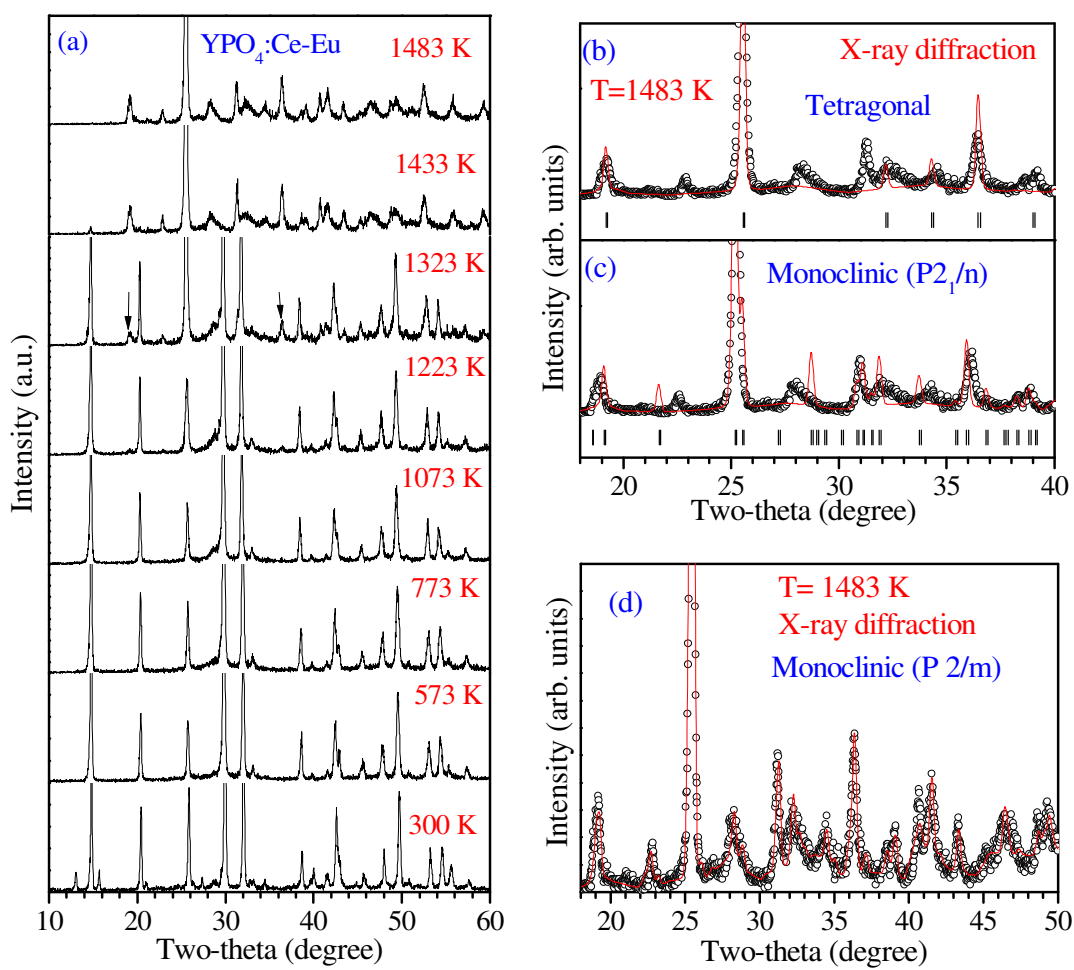

**Figure 8.** (a) Evolution of a portion of powder X-ray diffraction patterns of YPO$_4$:Ce-Eu at selected temperatures. Observed (circle) and calculated (continuous line) profiles obtained after the Rietveld refinement (structure-free Lebail fitting) of YPO$_4$:Ce-Eu, using (b) tetragonal and (c& d) monoclinic phases at 1483 K.